\documentclass{emulateapj}
\usepackage{subfigure}

\slugcomment{}

\shorttitle{Magnetic Bright Points}
\shortauthors{P. J. Crockett et al.}

\begin{document}

\title{The Area Distribution of Solar Magnetic Bright Points}

\author{P. J. Crockett, M. Mathioudakis, D. B. Jess, S. Shelyag, F. P. Keenan}
\affil{Astrophysics Research Centre, School of Mathematics and Physics, Queen's University, Belfast, BT7~1NN, 
Northern Ireland, U.K.}

\and

\author{D. J. Christian}
\affil{Department of Physics and Astronomy, California State University, Northridge, CA 91330, U.S.A.}

\begin{abstract}
Magnetic Bright Points (MBPs) are among the smallest observable objects on the solar photosphere. A combination of G-band observations and numerical simulations is used to determine their area distribution. An automatic detection algorithm, employing 1-dimensional intensity profiling, is utilized to identify  these structures in the observed and simulated datasets. 
Both distributions peak at an area of $\approx$45000~km$^2$, with a sharp decrease towards smaller areas. The distributions conform with log-normal statistics, which suggests that flux fragmentation dominates over flux convergence. Radiative magneto-convection simulations indicate an independence in the MBP area distribution for differing magnetic flux densities. The most commonly occurring bright point size corresponds to the typical width of intergranular lanes.

\end{abstract}


\section{Introduction}
\label{intro}
The dominant pattern covering the quiet solar photosphere is the granulation. Plasma flows remove the magnetic flux from granules into the dark inter-granular lanes where it clusters together to form small magnetic concentrations of 1$^{\prime\prime}$ or less, with field strengths often in excess of a kilogauss. Magnetic Bright Points (MBPs) are a manifestation of these kilogauss fields. They are among the smallest observable objects on the photosphere appearing as intensity enhancements within the inter-granular lanes. They are formed by complex processes involving the interaction of magnetic fields with the convectively unstable plasma and can provide a conduit for channeling kinetic energy into the upper atmosphere \cite[]{deWijn09}. Although all MBPs are found to reside in areas of high magnetic field, the opposite is not always true, with some high magnetic field in plage regions appearing  devoid of MBPs \cite[]{Ish07}.  An examination of the physical parameters required for a MBP to exist, i.e. minimum or maximum size, and lower magnetic field threshold, will further our knowledge on the creation and evolution of such small-scale photospheric magnetic fields.

The physical processes responsible for MBP formation have been simulated by \citet{shelyag2a} and \citet{shelyag2}.  Magneto-convection models for convectively unstable plasma in the photosphere, are combined with the radiative properties of that plasma and include the partial ionization of hydrogen and the other most abundant elements. The validation of numerical simulations by observations, may allow us to conclude whether the physics in the simulations can describe the real Sun and hence help us interpret the observational findings.

G-band imaging forms a common basis for MBP observations. \cite{Alm04} find that MBPs have a maximum intensity 1.8 times the mean photospheric value. Several authors propose that such intensity enhancements are caused by a significant weakening of the CH absorption lines, due to the dissociation of CH molecules at increasing temperatures \cite[]{Alm01,Ste01,Lan02}.  The latter group identify two types of bright points that occur in G-band images; those that are associated with magnetic structures, and others that exist at the edge of rapidly expanding granules.  The latter are believed to be density enhancements, caused by a build-up of material attempting to flow down the thin inter-granular lanes at the perimeter of a granule.     

The sheer number of MBPs requires automated algorithms for their detection and tracking. In general, higher spatial resolution leads to smaller sizes being detected.  \cite{Ber95} applied an altered blob-finding algorithm to separate MBPs from granules. They quantified the MBP size in terms of the FWHM intensity diameter, taking the smallest dimension across the identified objects.  A semi-automatic process was implemented that included nonlinear least-squares Gaussian fitting to the observed intensity profiles. Their analysis revealed  a modal diameter of 220~km, an average diameter of 250~km, and a diameter range of 120--600~km. \cite{Bov03} employed a specialized version of the multiple-level tracking pattern recognition software \citep[][]{Bov01}, which applies several decreasing intensity thresholds to an image.  
A dominant diameter of 220 $\pm$ 25 km was detected.  \cite{Wie04} repeated this procedure using higher spatial resolution observations, and found a predominant diameter of 160 $\pm$ 20 km. \cite{Alm04} visually identified MBPs in individual G-band images. The area was determined by a segmentation algorithm, and their diameter measured by fitting a double gaussian profile. They obtain 135km for the minor axis, which seems to be set by the angular resolution of the observations. This double gaussian decontaminates the profiles, by taking into account the intensity distribution of the dark local background, within which the MBPs are situated \citep[][]{Tit96}. \cite{Utz09} utilized an altered version of the \cite{Bov01} algorithm, and applied it to Hinode SOT observations.  The size of MBPs was defined by placing an upper and lower intensity threshold on the segmented structures, resulting in mean diameters of 166-218~km.  

In this paper we use observations and numerical simulations to investigate the area distribution of MBPs. An automatic detection and tracking algorithm, described in \cite{Croc09} (hereafter, Paper 1), is developed further and applied to high resolution G-band images. 
Section 2 discusses the observations, with emphasis on an automated algorithm used for MBP detection and size determination.  A description of the numerical simulations are given in Section 3. Our main findings are presented in Section 4, with concluding remarks in Section 5.

\section{Observations and Data Analysis}
\label{data}    
\begin{figure}
\epsscale{1.0}
\plotone{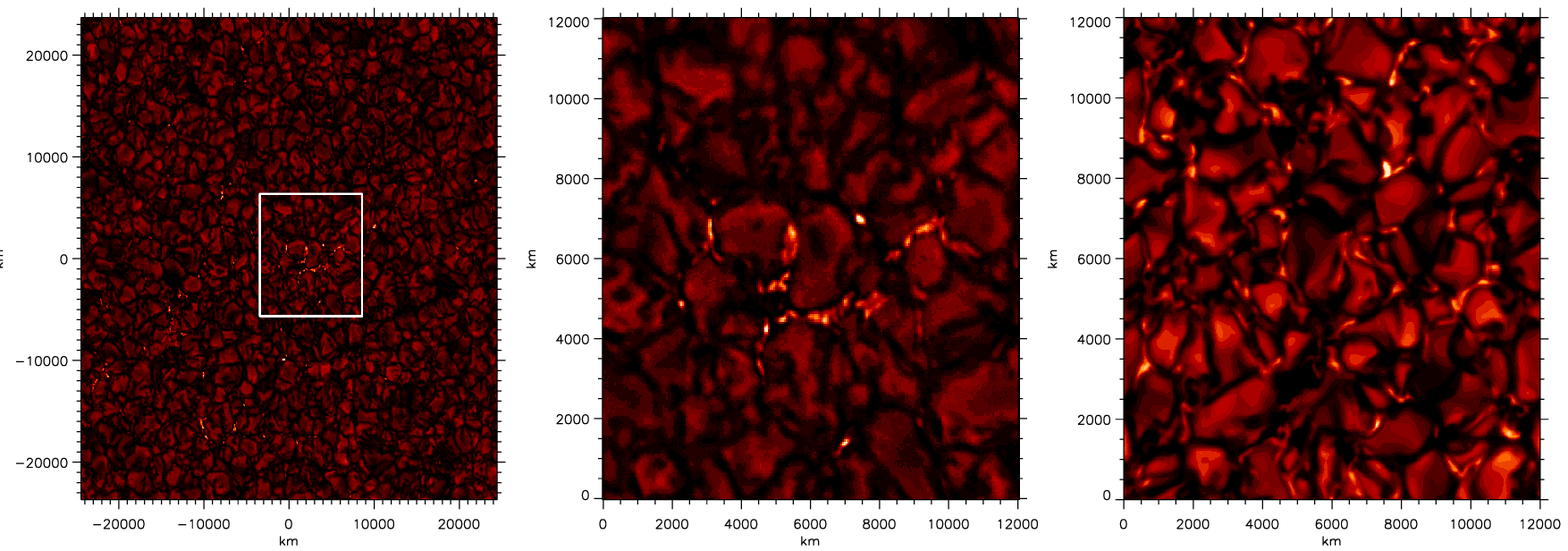}
\caption{{\it{Left:}} A $70'' \times 70''$ G-band image from the ROSA dataset. {\it{Middle:}} A $12 \times 12$~Mm section of the G-band image. {\it{Right:}} G-band simulation for an average field of 200G.
\label{f1}}
\end{figure}

The data was obtained by the newly-commissioned Rapid Oscillations in the Solar Atmosphere (ROSA) instrument, installed at the 76~cm Dunn Solar Telescope (DST), in New Mexico, USA \citep[]{Jess10}. The observations were taken on 28 May 2009 through a 12{\AA} filter, centered at 4305{\AA} (G-band), during a period of excellent seeing. Post-facto speckle reconstruction algorithms \cite[]{Wog08}, in addition to rigorous image de-stretching using a $40 \times 40$ grid, (equating to a $\approx$1.7$''$ separation between spatial samples, \citep{Jes08}), was implemented to remove effects caused by atmospheric seeing. We observed a $70'' \times 70''$ quiet solar region at disk center, achieving diffraction-limited imaging with 0.069$''$ pixel$^{-1}$. Figure~\ref{f1} displays a typical G-band image from the dataset, with multiple MBPs visible in the central region.
Analysis of the data was performed with an updated version of a detection algorithm described in Paper 1, which uses intensity thresholding to map the intergranular lanes. A compass search allows MBPs to be disentangled from bright pixels within granules, while object growing accounts for any pixels that might have been removed when mapping the lanes. One of the disadvantages of the algorithm described in Paper 1 is the requirement for the image to be divided into segments, with each subsequent segment being processed individually. Here we use an updated algorithm which operates on the entire $70'' \times 70''$ image sequence, thus improving computational time and accuracy.  This development is particularly important, as it permits accurate estimates of MBP areas. 
Mapping the location of the inter-granular lanes, with an overestimation of the intensity threshold, is used to separate out bright objects.  The threshold set is the mean image intensity plus one sigma. All structures under this level are considered a lane and are not investigated by our algorithm.  The vast majority of MBPs retain higher intensities, however some very dull MBPs may be lost at this stage.  Objects are then investigated individually. We impose a 3-sigma intensity variation limit on each object, in order to fully separate MBPs from the granules.  Any bright object possessing an intensity range greater than 3 sigma is broken up into smaller objects, until the resulting structures comply with this condition.

\begin{figure}
\epsscale{0.6}
\plotone{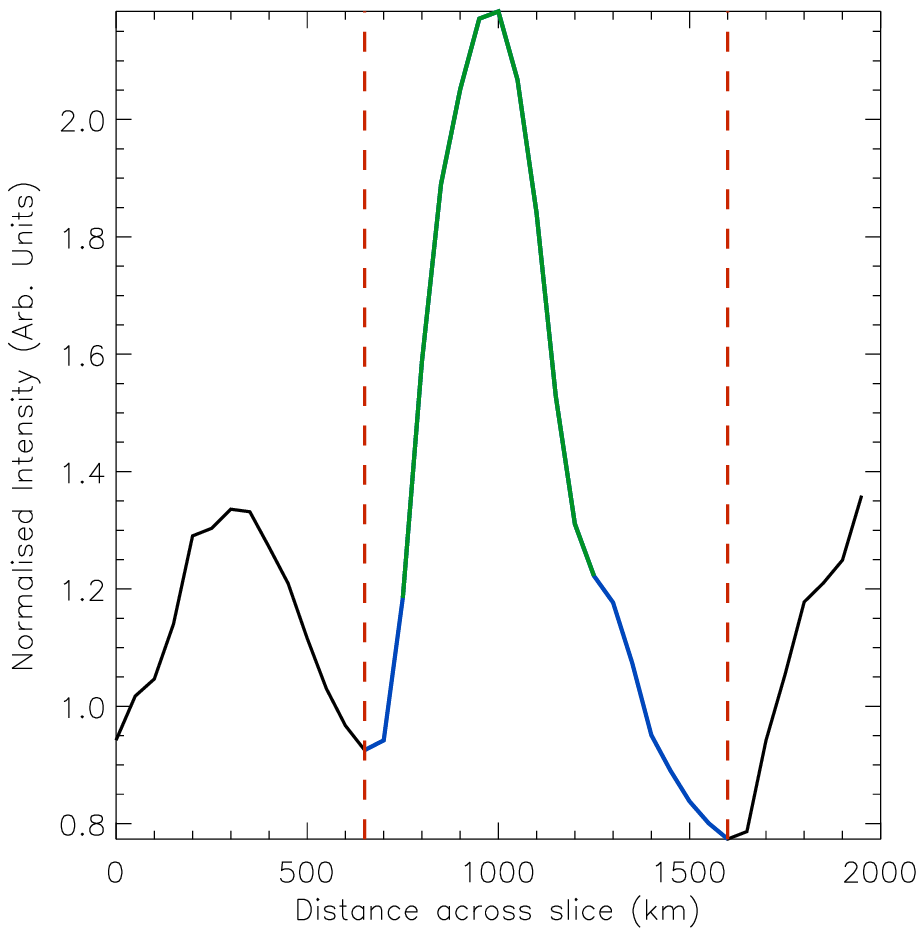}
\plotone{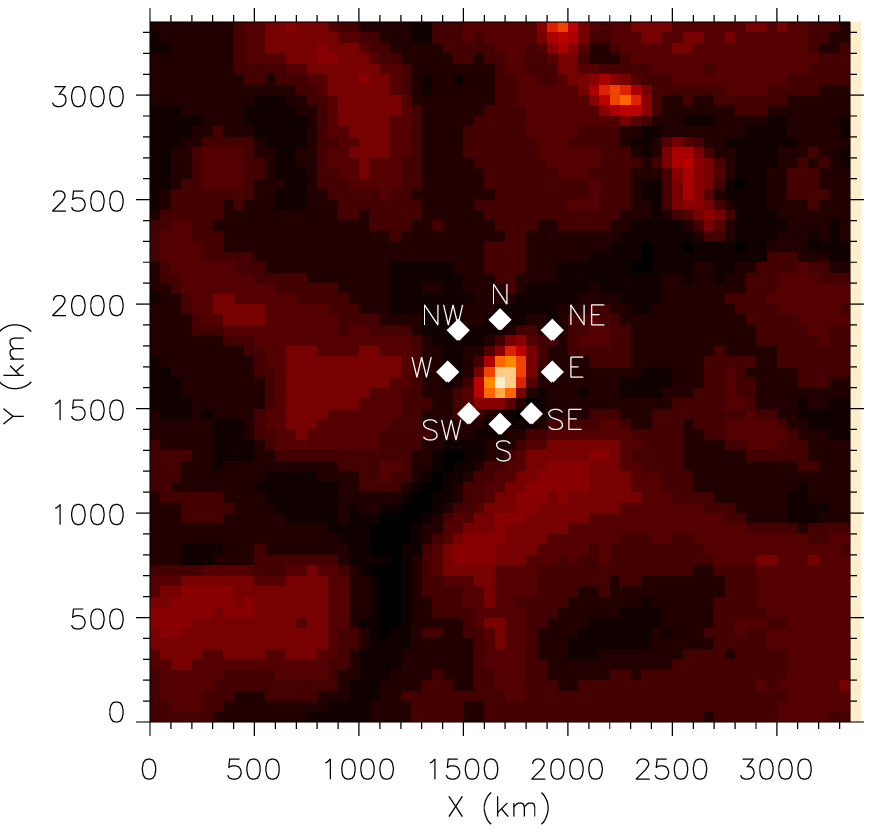}
\caption{{\it{Top:}} Intensity profile typical of a MBP. A steep intensity gradient is followed by two minimum turning points at the centre of the dark lane. The algorithm detects the minimum turning points, marked by dashed red lines, in 8 separate directions. {\it{Bottom:}} The white marks indicate the locations of turning points in the image. These turning points reside in the center of the inter-granular lanes. The MBP shown has an area of 127,500 km$^{2}$.
\label{f2}}
\end{figure}

The detection of MBPs is carried out by an extended version of the compass search (see Paper 1 {\S}4.3), and incorporates gradient thresholding through intensity profiling.  A one-dimensional variation in intensity, across a selected region of the image, is first determined (see left panel of Fig.~\ref{f2}).  Intensity profiles for each object are established for eight directions, symmetrically positioned about the objects centre-of-gravity.  The stipulation that a lane must be in close proximity to the MBP remains.  The algorithm now actively searches for inter-granular lanes, by using the turning points of the intensity profiles,  which are located at the centre of the lanes (left panel of Fig.~\ref{f2}).  Hence, the lanes are located from in-situ intensity profiling. Each measurement is specific, not only to individual objects, but in every considered direction as well.  To establish each turning point the one-dimensional line, from which intensity profiles are procured, is extended until two stationary points exist in the profile, i.e. where the rate of change in intensity (y) as a function of distance (x) is zero, dy/dx = 0.  A limit on the distance between the turning points eliminates large objects, such as granules.

Gradient thresholding is applied to all intensity profiles to disentangle MBPs from granules.  MBPs retain a very steep intensity change in all directions, compared to a more gradual variation associated with granules. The maximum gradient is determined from any part of a profile falling between the two turning points (see Fig.~\ref{f2}).  The threshold gradient is derived for each individual image through a selection of 500 random objects, and is calculated prior to the execution of the algorithm.  A threshold is determined by adding a 1-sigma value to the median gradient recorded for each image.

\begin{figure}
\begin{minipage}[t]{0.5\linewidth}
\epsscale{1.0}
\plotone{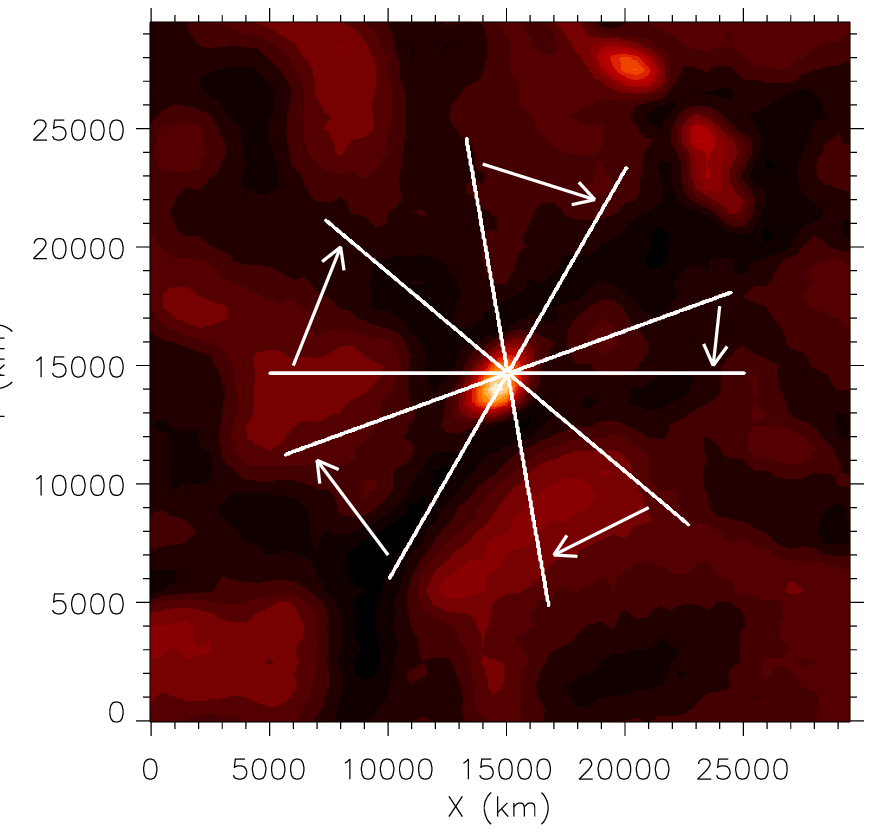}
\end{minipage}
\hspace{-1.2cm}
\begin{minipage}[t]{0.5\linewidth}
\epsscale{1.0}
\plotone{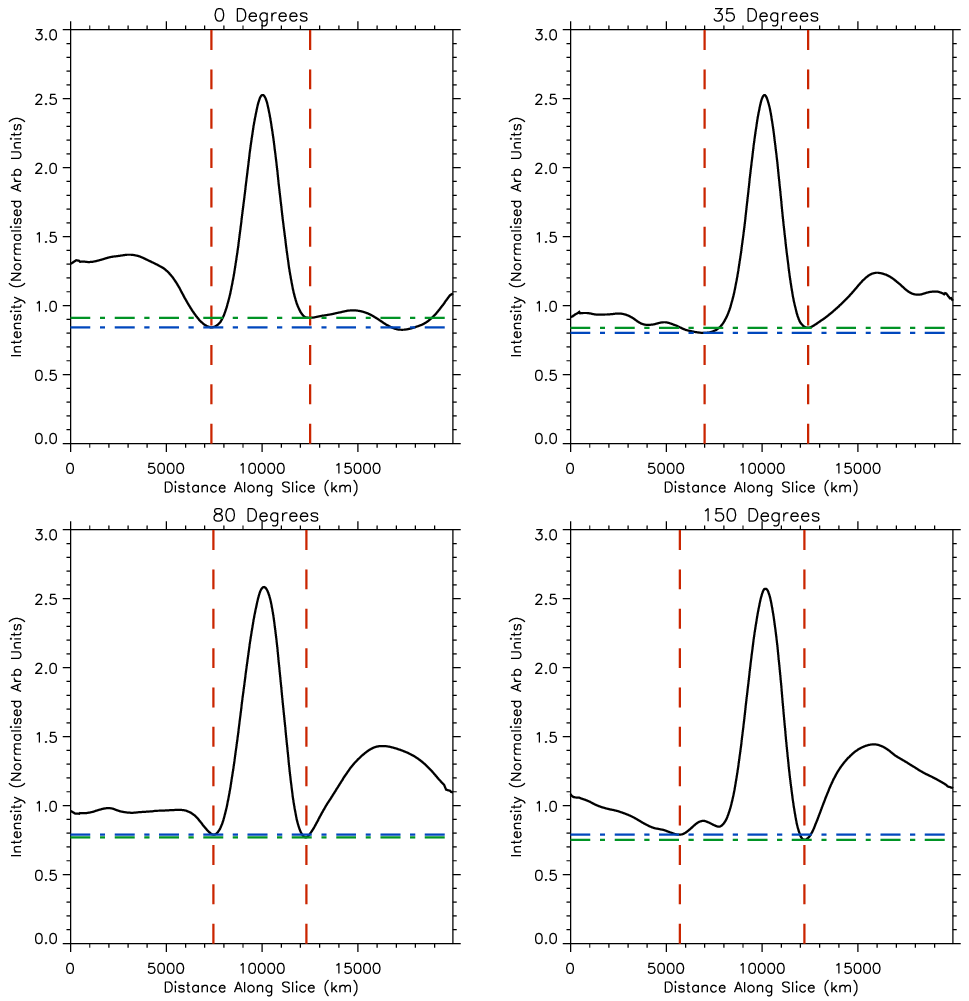}
\end{minipage}
\caption{A line is rotated, in 5 degree steps, around an object to create one-dimensional intensity profiles (right). The turning points are marked by dashed red lines, whilst the green and blue dot-dash lines indicate the intensity levels of the left and right turning points, respectively.  The maximum turning point, across all profiles, is imposed as the lower threshold for MBP growing. Note the re-binning of the data to ensure turning points are located accurately.
\label{f3}}
\end{figure}

A significant improvement of the present algorithm concerns the growing of MBPs. A newly developed process provides an independent threshold range for each object for accurate area representation.  The algorithm rotates a one-dimensional line through 360 degrees, in 5 degree steps, about an object's centre of gravity.  Intensity values at the turning points of the profiles, i.e. the lanes, are acquired at each angle.  To aid the accurate determination of turning points, the data is re-binned by a factor of ten  and smoothed.  Thus a narrow intergranular lane and the associated turning point can be clearly identified.  The maximum turning point intensity is taken to provide a lower cutoff to our growing algorithm, while the upper boundary is set as the maximum intensity level occurring within the seed region.    The growing procedure includes any conjoining pixels that are above the lower threshold cutoff. The MBP area is determined by totaling the number of pixels within each structure. Our sampling of 0.069$''$ pixel$^{-1}$ provides an area of 2500~km$^2$pixel$^{-1}$.  
This procedure, demonstrated in Figure~\ref{f3}, reproduces 90\% of MBPs to within a 10\% error of visually identified areas.  Setting the lower intensity threshold as the brightest surrounding lane enforces an upper limit on the area of the MBPs, i.e MBPs are grown to their maximum dimensions.


\section{Numerical Simulations}
\label{simulations}

We use the MURaM code \citep{shelyag1} to carry out simulations of radiative magneto-convection in the upper solar convection zone and photosphere. This code uses a fourth-order, central difference scheme for computing the spatial derivatives, and a fourth-order, Runge-Kutta scheme to advance the solution forward in time. The solution is stabilized against numerical instabilities using additional artificial hyperdiffusive terms, described in detail by \cite{caunt}, \cite{shelyag1} and \cite{shelyag3}. The size of the computational domain used for the simulations is $12 \times 12 \times 1.4$~Mm$^{3}$, resolved by $480 \times 480 \times 100$ grid cells providing a resolution of 25 km per grid cell. However, we emphasize that as a result of hyperdiffusivity, the size of the smallest structures produced in the smulations can be larger than a single grid cell.  Due to the dependence of the hyperdiffusivity coefficients on the local solution, it is not possible to globally define a quantity, uniquely representing the resolution.  However, a standard test, such as the strong (compression ratio 100) Riemann shock tube \cite[]{Sod78}, can be used to provide an indication of the resolution. The results of such tests for similar codes (i.e. 4-6-th order central difference spatial scheme and hyperdiffusive sources) show that even for such an extreme  case, the shock front is diffused over 2-4 grid cells, depending on the relative position of the shock front with respect to the grid (Caunt \& Korpi, 2001; Shelyag et al., 2008). Consequently, the resolution of the code for this case is about 50-100 km, a value similar to the resolution of the observations. The side boundaries are periodic, the upper boundary is closed for vertical and stress-free horizontal plasma motions, while the bottom boundary is transparent. The level corresponding to the visible solar surface is located approximately 400~km below the upper boundary. This setup allows us to perform radiative diagnostics of G-band images, and directly compare them with the observations. A detailed description of the method used is given in \cite{shelyag2}. Here we provide a brief description of the process. 

For each of the light rays corresponding to a vertical plasma column in the simulation, we compute the LTE spectrum in the 4295--4315~{\AA} range, which consists of 328 absorption lines, 239 of which are produced by CH molecules. The calculated spectrum is convolved with the  G-band filter function. The magnetic splitting of CH lines, and its influence on G-band intensities, are sufficiently small for this effect to be neglected. G-band images obtained using this technique reproduce the dynamic and radiative properties of magneto-convection, and show a large number of G-band bright points, corresponding to the heated, and partially evacuated, magnetic flux tubes seen in Figure~\ref{f1}.

\section{Results}
\label{results}
\begin{figure}
\epsscale{1.0}
\plotone{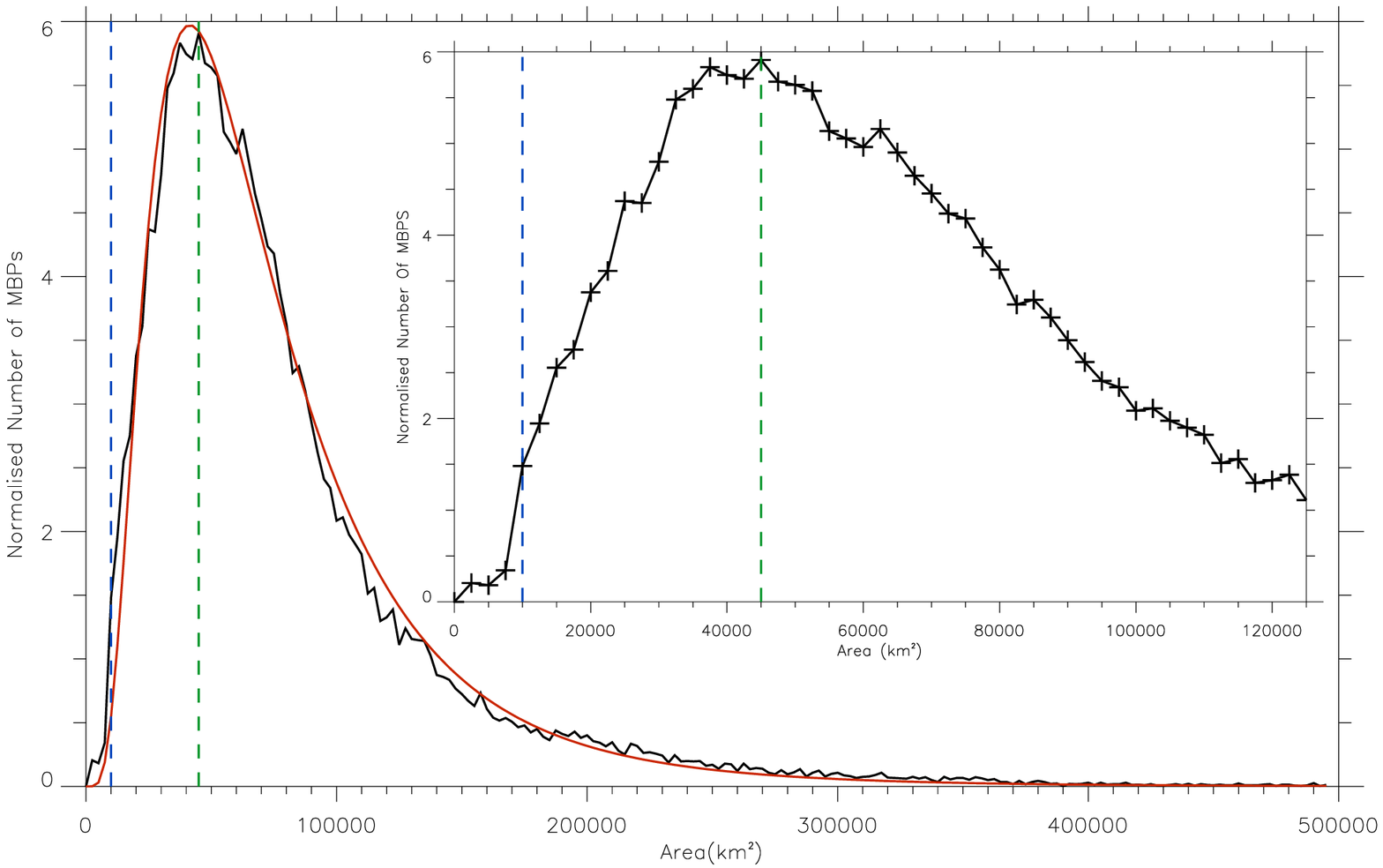}
\plotone{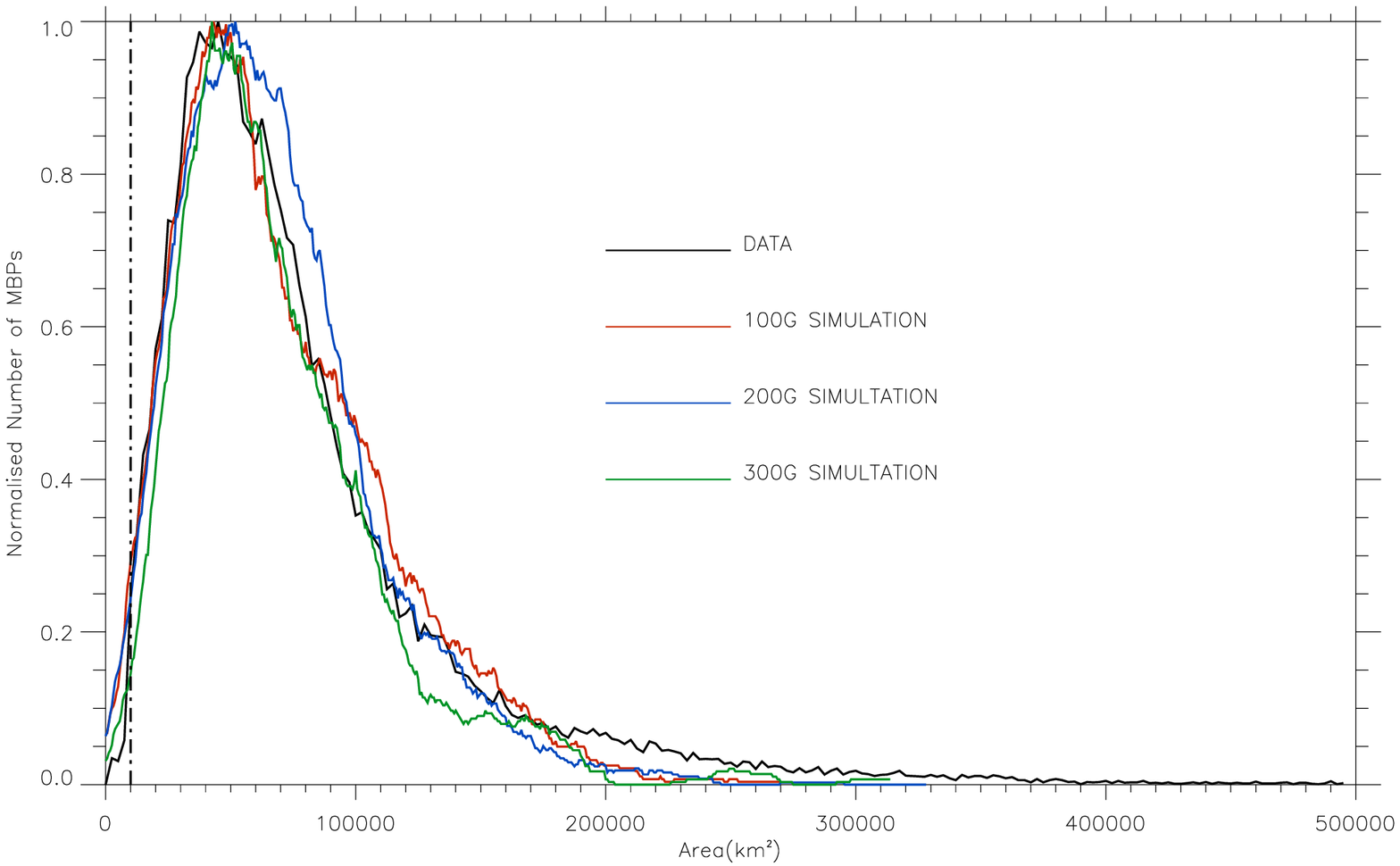}
\caption{{\it{Top:}} The observed area distribution of MPBs. The green and blue dashed lines mark, respectively, the peak of the distribution at 45000~km$^2$, and the diffraction limit at 10000~km$^2$. A log-normal fit to the distribution is overplotted as a red line. {\it{Insert:}} Expanded plot of the distribution around the peak. {\it{Bottom:}} The observed distribution is compared to simulations derived for average magnetic fields of 100G, 200G and 300G respectively. 
\label{f4}}
\end{figure}

A series of 500 images were investigated, incorporating a total of 63312 MBPs.  The MBPs cover approximately 0.42\% of the solar surface with a variance between 0.33\% and 0.53\% across the time series. 
Figure~\ref{f4} displays the area distribution of MBPs, with their occurrence normalized to the mean number detected across all bins. The distribution was created by a summation of MBPs across all images, in 1~pixel bins.  This technique may lead to the ``double counting'' of MBPs, some of which may have longer lifetimes than others.  However, snapshots of single frames produce a similar distribution, with approximately the same peak.  Therefore, this technique is equivalent to the summation of multiple snapshot distributions, each with similar parameters, resulting in an overall identical distribution. We therefore believe that any double counting does not pose a problem in our interpretation.

The nature of the distribution appears to conform with log-normal statistics. To confirm this, a log-normal probability density function (PDF) of the form, 
\begin{equation}
PDF_{log-normal}=\frac{1}{x\mu\sqrt{2\pi}}\exp\frac{-(\ln x - \mu)^2}{2\sigma^2} \ ,
\end{equation} 
where $\mu$ and $\sigma$ are, respectively, the mean and standard deviation of $\ln$ x, is fitted to the data.  Values of $\mu$=3.25 and $\sigma$=0.65 produce an excellent fit, shown by the over-plotted solid red line in Figure~\ref{f4}.  To quantify the goodness of the fit, the $\chi^2$ error statistic \cite[]{Wal03} of the form,  
\begin{equation}
\chi^2 = \sum_{i=1}^{n} \frac{(O_i - E_i)^2}{E_i}
\end{equation}
where $O_{i}$ and $E_{i}$ are, respectively, the observed and expected frequencies, is utilized.  The observed frequencies correspond to the values obtained in the data, whilst the expected frequencies correspond to the theoretical values set by the log-normal fit.  
Comparison of the real data with the fitted distribution reveals a conformity of 99.5\%, and confirms the MBP area distribution is well described by  log-normal statistics.

The peak of the distribution occurs at an area of 45000~km$^{2}$.  Assuming a circular geometry, this corresponds to a diameter of 230~km.  
While this estimate appears in general agreement with earlier works, there are a number of points that need to be emphasized.\cite{Utz09} find diameters of 218$\pm$48~km using a spatial sampling of 0.108$''$ pixel$^{-1}$ on Hinode SOT but this is dependent on the spatial sampling. Reducing the spatial sampling to 0.054$''$ pixel$^{-1}$ gives a diameter of 166$\pm$31~km. The latter value is in agreement with the results of \cite{Wie04}.  Differences in the diameters may be explained by differences in the detection algorithms employed.
\cite{Wie04} employ the Multi Level Tracking (MLT) algorithm which utilizes  decreasing intensity levels to identify and separate objects. MLT sets an initial uppermost intensity level. Bright structures which exceed this threshold are tagged. The intensity level is then lowered. Pixels above the new intensity level, adjacent to the structures identified in the previous level, are added. New structures that appear at this level are tagged separately. This repetitive procedure is terminated after a last extension to a final intensity level deemed adequate for representing the observed pattern. The structures are forced to be separated by 2 pixels on all sides. The enforced separation and the somewhat arbitrary final threshold level, may affect the dimensions of the MBPs measured by missing dim edge pixels. \cite{Utz09} employ a similar repetitive intensity thresholding technique to separate granules and MBPs. They impose an upper and lower intensity boundary on the pixels of an object to determine its size. The upper boundary is given by the maximum intensity in the object whilst the lower boundary is defined as the maximum minus 30\% of the mean photospheric intensity.  Again these conditions may limit the final size of MBPs to the brightest pixels. This effect may be exaggerated at the higher spatial sampling.   
Instead our algorithm sets a threshold which is specific to each MBP. Moreover by taking the threshold as the highest intensity within the lane, we determine an upper limit in the MBP area 
including dim edge pixels which may be neglected in the earlier studies.

In Figure~\ref{f4}, the vertical dashed line at 10000~km$^{2}$ marks a 4-pixel (2$\times$2) resolution threshold, thus placing a limit on the smallest structures that can be resolved in our observations. The procedure of identifying the smallest structures in the observations was tested by convolving the simulated G-band images with the Airy function corresponding to a 76 cm aperture and rebinning them to the spatial sampling of the observations. 
This test showed that 11 out of 12 MBPs, each covering an area of 3-5 pixels in the degraded images, correspond to MBPs in the original (non-degraded) images.  
The sharp drop in the area of MBPs below the peak, implies that underlying physical processes are limiting the creation, and evolution, of very small  structures.
In addition, the distribution shows only a relatively small number of large-scale MBPs with area greater than 200000~km$^2$.  

Errors in the distribution were determined by comparing the visual estimates of the area with the output of the algorithm. The comparison reveals the  distribution of errors for each bin from which a sigma value was determined. At the peak of our distribution the error is $\pm$8$\%$ ($\pm$3600 km). The uncertainty at the high end of the distribution decreases whilst, as we approach the diffraction limit the errors become larger due to small number statistics.

The simulations allow us to study the effect of changing the net magnetic flux density on solar granulation and how this modifies the properties of MBPs. To produce a theoretical distribution of their area, we first computed a series of consecutive G-band snapshots, based upon different average magnetic flux densities. The same procedures for MBP detection and area estimation were then applied directly to these simulations. We utilized 100, 200, and 300~G signed vertical magnetic flux simulations, and compared the resulting distributions directly with the observations. The spatial resolution of simulations has not been degraded (Fig ~\ref{f4}). All three distributions, based on different initial magnetic flux densities, agree with the log-normal form of the observed MBP area distribution. Crucially, these simulated G-band images also exhibit the same sharp decline from the peak to the diffraction limit. The peak of the distributions do not alter significantly, with a simulated maximum at $\approx$50000~km$^{2}$.  The simulations were degraded by re-binning the data to match the observational scale of 50km per pixel.  A further comparison revealed that the simulated distribution remained unchanged.    


\section{Concluding Remarks}
\label{conc}

We utilize an improved automatic detection algorithm to study the area distribution of solar magnetic bright points (MBPs). 
The area distribution of MBPs follows log-normal statistics. We interpret this result in a similar fashion as \cite{Bog88}, who suggest that the  underlying fragmentation process is responsible for log-normal distributions.  Similarly, we can not rule out coalescence.




The peak of our MBP distribution occurs at $45000~km^{2}$ significantly higher than our telescope diffraction limit, and is consistent with the results of the radiative MHD simulations. 
The minimal area of MBPs is most likely defined by the width of the intergranular lanes, which is subsequently limited by the radiative and convective energy balance, and mass conservation in magneto-convective processes. The peak in the area distribution seems to correspond to the most probable width of the intergranular lane. The area of large MBPs may be limited by the lack of sufficient radiative heating in the larger flux tubes. As has been demonstrated \citep[see e.g.][]{Ber95}, wall heating of magnetic flux concentrations is not sufficient to increase the vertical radiative flux for flux tubes greater than 500~km in width. Thus, MBPs cannot be generated in large diameter magnetic flux tubes. Elongated bright points of large size may still, in principle, be formed. However, their formation will be inhibited by strong plasma motions and granule fragmentation.



{\bf Acknowledgments}\\
This work has been supported by the UK Science and Technology Facilities Council (STFC). Observations were obtained at the National Solar Observatory, operated by the Association of Universities for Research in Astronomy, Inc (AURA), under cooperative agreement with the National Science Foundation.  PJC thanks the Northern Ireland Department for Employment and Learning for a PhD studentship.  DBJ is grateful to STFC for the award of a post-doctoral fellowship. FPK is grateful to AWE Aldermaston for the award of a William Penney Fellowship. We wish to thank the anonymous referee for useful comments and suggestions.

\newpage

\clearpage

\newpage

\begin{thebibliography}{}
\bibitem[Bogdan et al.(1988)]{Bog88} 
Bogdan, T.~J., Gilman, P.~A., Lerche, I., \& Howard, R.\ 1988, \apj, 327, 451 
\bibitem[Berger et al.(1995)]{Ber95} 
Berger, T.~E., Schrijver, C.~J., Shine, R.~A., Tarbell, T.~D., Title, A.~M., \& Scharmer, G.\ 1995, \apj, 454, 531 
\bibitem[Bovelet \& Wiehr(2001)]{Bov01} 
Bovelet, B., \& Wiehr, E.\ 2001, \solphys, 201, 13 
\bibitem[Bovelet \& Wiehr(2003)]{Bov03} 
Bovelet, B., \& Wiehr, E.\ 2003, \aap, 412, 249 
\bibitem[Caunt \& Korpi(2001)]{caunt} 
Caunt, S.~E., \& Korpi, M.~J.\ 2001, \aap, 369, 706 
\bibitem[Crockett et al.(2009)]{Croc09} 
Crockett, P.~J., Jess, D.~B., Mathioudakis, M., \& Keenan, F.~P.\ 2009, \mnras, 397, 1852 
\bibitem[de Wijn et al. (2009)]{deWijn09}
de Wijn, A.G., Stenflo, J.O., Solanki, S.K., \& Tsuneta, S.\ 2009, Space Science Reviews, 144,
275  
\bibitem[Ishikawa et al.(2007)]{Ish07} 
Ishikawa, R., et al.\ 2007, \aap, 472, 911
\bibitem[Jess et~al.(2010)]{Jess10}
Jess, D.~B., Mathioudakis, M., Christian, D.~J., Keenan, F.~P., Ryans, R.~S.~I., \& Crockett, P.~J.\ 2010, \solphys, 261, 363 
\bibitem[Jess et~al.(2008)]{Jes08}
Jess, D.~B., Mathioudakis, M., Crockett, P.~J., \& Keenan, F.~P.\ 2008, \apjl, 688, L119 
\bibitem[Langhans et al.(2002)]{Lan02} 
Langhans, K., Schmidt, W., \& Tritschler, A.\ 2002, Solar Variability: From Core to Outer Frontiers, 506, 455
\bibitem[S{\'a}nchez Almeida et al.(2001)]{Alm01} 
S{\'a}nchez Almeida, J., Asensio Ramos, A., Trujillo Bueno, J., \& Cernicharo, J.\ 2001, \apj, 555, 978 
\bibitem[S{\'a}nchez Almeida et al.(2004)]{Alm04} 
S{\'a}nchez Almeida, J., M{\'a}rquez, I., Bonet, J.~A., Dom{\'{\i}}nguez Cerde{\~n}a, I., \& Muller, R.\ 2004, \apjl, 609, L91 
\bibitem[Sch{\"u}ssler et al.(2003)]{shelyag2a} 
Sch{\"u}ssler, M., Shelyag, S., Berdyugina, S., V{\"o}gler, A., \& Solanki, S.~K.\ 2003, \apjl, 597, L173 
\bibitem[Shelyag et al.(2004)]{shelyag2} 
Shelyag, S., Sch{\"u}ssler, M., Solanki, S.~K., Berdyugina, S.~V., V{\"o}gler, A.\ 2004, \aap, 427, 335 
\bibitem[Shelyag et al.(2008)]{shelyag3} 
Shelyag, S., Fedun, V., \& Erd{\'e}lyi, R.\ 2008, \aap, 486, 655 
\bibitem[Sod(1978)]{Sod78} 
Sod, G.~A.\ 1978, Journal of Computational Physics, 27, 1 
\bibitem[Steiner et al.(2001)]{Ste01} 
Steiner, O., Hauschildt, P.~H., \& Bruls, J.\ 2001, \aap, 372, L13 
\bibitem[Title \& Berger(1996)]{Tit96} 
Title, A.~M., \& Berger, T.~E.\ 1996, \apj, 463, 797
\bibitem[Utz et al.(2009)]{Utz09} 
Utz, D., Hanslmeier, A., M{\"o}stl, C., Muller, R., Veronig, A., \& Muthsam, H.\ 2009, \aap, 498, 289 
\bibitem[V{\"o}gler et al.(2005)]{shelyag1}
 V{\"o}gler, A., Shelyag, S., Sch{\"u}ssler, M., Cattaneo, F., Emonet, T., \& Linde, T.\ 2005, \aap, 429, 335 
\bibitem[Wall \& Jenkins(2003)]{Wal03} 
Wall, J.~V., \& Jenkins, C.~R.\ 2003, Princeton Series in Astrophysics, Cambridge University Press
\bibitem[W{\"o}ger et al. (2008)]{Wog08} 
W{\"o}ger, F., von der L{\"u}he, O., Reardon, K.\ 2008, \aap, 488, 375
\bibitem[Wiehr et al.(2004)]{Wie04}
Wiehr, E., Bovelet, B., \& Hirzberger, J.\ 2004, \aap, 422, L63 
\end{thebibliography}
\end{document}